\documentclass[a4paper,english]{paper} 

\usepackage[english]{babel}
\usepackage[babel,english=american]{csquotes}
\usepackage[margin=2.2cm]{geometry}
\usepackage{graphicx}

\usepackage{amsmath}
\usepackage{upgreek}
\usepackage[font=small,labelfont=bf,]{caption}
\usepackage[format=plain,indention=0cm,labelfont={bf}]{caption}[2006/03/21]

\usepackage[list=true]{subcaption}
\usepackage{natbib}
\usepackage{booktabs} 
\usepackage{xcolor}
\usepackage[bottom,perpage,hang,symbol*]{footmisc}
\usepackage[colorlinks=true,
			bookmarks=true, bookmarksopen=true, bookmarksopenlevel=1, bookmarksnumbered=true,
			citecolor=blue, linkcolor=blue, urlcolor=blue,
			pdftitle={ECC - energy calibration via correlation},
			pdfauthor={Daniel Maier},
			pdfsubject={},
			pdfcreator={TeXnicCenter 2.02, (stable) 64 bit},
			pdfproducer={pdfTeX 3.1415926 (1.40.12) (MiKTeX 2.9)}
			]{hyperref}
\usepackage{hypcap}
\newcommand{\PHA}{\mathit{PHA}}
\newcommand{\Am}{$^{241}\!$Am}

\sectionfont{\large\sf\bfseries\color{black!70!blue}} 
\title{Energy calibration via correlation}
\subtitle{\normalsize{Originally published in Nuclear Instruments and Methods in Physics Research Section A, 2016 \\ (DOI:	10.1016/j.nima.2015.11.149)}}
\author{Daniel Maier, Olivier Limousin} 
\institution{CEA Saclay, DSM/Irfu/Service d'Astrophysique, 91191 Gif-sur-Yvette Cedex, France}

\begin{document}
\sloppy
\twocolumn[\vspace{-1.4cm}\maketitle 
\hrule 
\begin{abstract} 
The main task of an energy calibration is to find a relation between pulse-height values and the corresponding energies. Doing this for each pulse-height channel individually requires an elaborated input spectrum with an excellent counting statistics and a sophisticated data analysis.
This work presents an easy to handle energy calibration process which can operate reliably on calibration measurements with low counting statistics.
The method uses a parameter based model for the energy calibration and concludes on the optimal parameters of the model by finding the best correlation between the measured pulse-height spectrum and multiple synthetic pulse-height spectra which are constructed with different sets of calibration parameters. A CdTe-based semiconductor detector and the line emissions of an \Am~source were used to test the performance of the correlation method in terms of systematic calibration errors for different counting statistics. Up to energies of 60\,keV systematic errors were measured to be less than $\sim\!0.1\,$keV. 
Energy calibration via correlation can be applied to any kind of calibration spectra and shows a robust behavior at low counting statistics. It enables a fast and accurate calibration that can be used to monitor the spectroscopic properties of a detector system in near realtime.
\end{abstract}
\vspace{-2mm}
\hrule\bigskip
]

\section{Introduction}

After digitizing detector signals, the pulse-height a\-nal\-y\-sis (PHA) of signals are expressed in the abstract analogue-to-digital unit (ADU) which is the binary output of the analogue-to-digital converter (ADC). In principle, knowing the exact response of the detector and of the preprocessing electronics (amplifier, pulse shaper, ADC) allows deriving the average energy of the absorbed radiation for each pulse-height value. Instead of this theoretical approach, a calibration which is based on a spectral measurement of a known calibration source can be applied for the transformation from PHA values to energies without knowing the detailed response of the detector system.

Using an ADC with an $N$-bit resolution, a \textit{full energy calibration} assigns for each of the $2^N$ pulse-height values a corresponding energy. 
Furthermore, a full calibration aims at measuring the quantum efficiency\footnote{Obtaining the quantum efficiency is not the goal of this work, but see the discussion in Section~\ref{sec:disc} on this topic.} of the detector system precisely. See \cite{Briel1999} for an overview of the extensive calibration test for the pn-CCD onboard the X-ray satellite XMM-Newton.

The high requirements concerning the calibration source and the large effort for the analysis of the calibration itself make a full energy calibration often not the first choice for detector systems which are in a development phase and which require a frequent recalibration due to changes of their system parameters like operating temperatures or voltage settings.

\subsubsection*{The line fitting approach}
\vspace{1.2mm}
Instead of a full energy calibration, discrete energy-pulse-height reference points (EPRPs) can be obtained by measuring the line emission of radio-iso\-topes or X-ray fluorescence lines. Fitting a parametric model according to the expected response of the detector system to these EPRPs allows deriving the optimal parameters for the energy calibration; see \cite{Majewski2014} for such an emission line fitting with a linear model and \cite{Jakubek2011} for a fit using a surrogate function. Because of uncertainties in the measurements of the EPRPs, a model with $p$ free parameters requires the determination of $q$ EPRPs with $q \geq p$. The reference points can be obtained by fitting $q$ Gaussian line profiles\footnote{In the following, a Gaussian shaped detector response is assumed.} to a calibration spectrum which is observed using $q$ monoenergetic emission lines.

\begin{figure*}[t]
\capstart
\centering
\includegraphics[width = 0.95\linewidth]{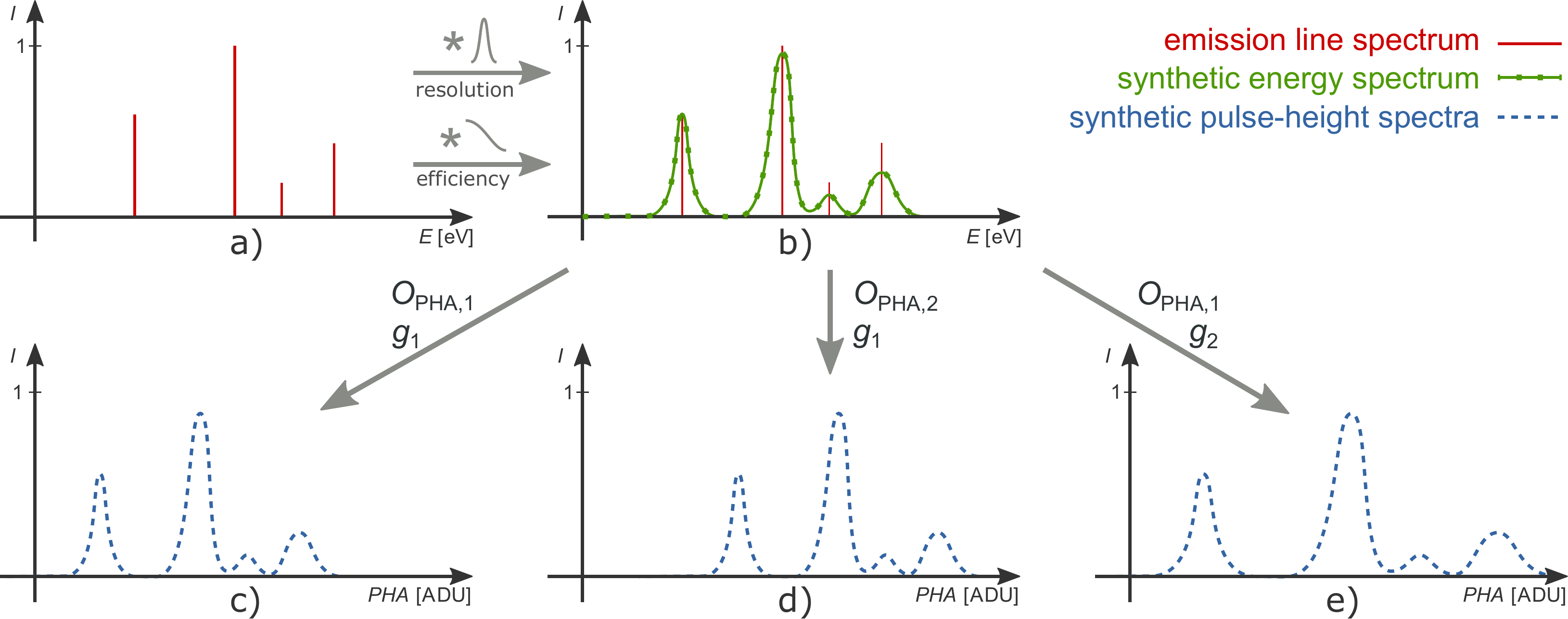}
\caption{Creation of synthetic pulse-height spectra. 
Top: the emission line spectrum [a),~red], normalized to the intensity of the most dominant line, is convolved with the expected energy resolution and quantum efficiency of the detector in order to create the synthetic energy spectrum [b), green]. Both, the energy resolution and the quantum efficiency are a function of the energy.
Bottom: using different parametrizations for the energy-to-pulse-height relation results in different synthetic pulse-height spectra. For a linear model, like it is shown, different offsets~$O_\mathrm{PHA}$ shift the spectra to lower or higher pulse heights; see the shift between the spectra (c) and (d) which are constructed with the same gain but different offset values. Different gain values $g$ stretch or squeeze the spectra which is shown for the spectra (c) and (e) which use the same offset but different gain values.}
\label{fig:ECC}
\end{figure*}

Additional to these requirements, the calibration source must have a sufficiently high flux. Detector systems consisting of $m$ independent channels, each with a slightly different response caused by inhomogeneities during their fabrication, require a channel specific energy calibration \cite{Majewski2014,Jakubek2011,Youn2014}. Instead of one calibration spectrum, $m$ spectra with sufficient statistics (more than 500 counts per channel are reported in \cite{Majewski2014}) must be obtained during the calibration measurement. To conclude, a precise energy calibration which is based on \textit{emission line fitting} puts high requirements on the calibration source and on the analysis of the channel specific calibration.

The theoretical basis of an energy calibration which is based on correlation (ECC)\footnote{Energy Calibration via Correlation.} is described in the following section. The core of the ECC method is similar to a fitting approach but uses all PHA values---and not just the peak positions---which results in an enhanced statistics of the ECC technique compared to a line fitting approach.

Section~\ref{sec:app} present the application of an ECC on the example of a linear and a non-linear calibration of a CdTe-based semiconductor detector system using the radio-isotope \Am. The spectroscopic implications are finally presented in Sect.\,\ref{sec:res}. Section~\ref{sec:perf} compares the calibration errors resulting from an ECC with the errors resulting from a line fitting approach for different counting statistics. The following discussion in Sect.\,\ref{sec:disc} is directed to other kinds of calibration sources and to the limitations of the ECC technique.

\section{Description of the method}
\label{sec:description}
\vspace{0.6mm}
Starting point of the ECC method is the known energy spectrum of the calibration source which is in the following called the \textit{synthetic energy spectrum} $I_\mathrm{syn}(E)$. The transformation between pulse-height values~$\PHA$ and energies~$E$ is defined via the parametric description $E(\PHA\,|\,P)$ or its inverse function $\PHA(E\,|\,P)$. Here and in the following, $f(A\,|\,B)$ describes a function $f$ with variable set $A$ and parameter set $B$. 
Different parameter sets $P$ result in different \textit{synthetic pulse-height spectra} $I_\mathrm{syn}(\PHA\,|\,P)$
\vspace{-1.0mm}
\begin{equation}
   I_\mathrm{syn}(\PHA\,|\,P) = I_\mathrm{syn}(E(\PHA\,|\,P)).
\end{equation}
The degree of correlation between a synthetic and the \textit{observed pulse-height spectrum} $I_\mathrm{obs}$ is obtained via a correlation factor $C$
\vspace{-1.0mm}
\begin{equation}
	\label{eq:cor}
	C(P) = \sum\limits_{\PHA=0}^{2^N-1} I_\mathrm{syn}(\PHA\,|\,P) \cdot I_\mathrm{obs}(\PHA).
\end{equation}
The spectral intensities $I_\mathrm{syn}$ and $I_\mathrm{obs}$ are normalized relative to the strongest line emission.
The parameter set $P^*$ with a correlation factor $C(P^*)$ that approaches the autocorrelation
\vspace{-2mm}
\begin{equation}
   \label{eq:c_max}
	C_\mathrm{max} = \sum\limits_{\PHA=0}^{2^N-1} I_\mathrm{obs}(\PHA) \cdot I_\mathrm{obs}(\PHA)
\end{equation}
is taken as the optimal parameter set for the transition between pulse heights and energies and can be used to transform all measured data via $E(\PHA\,|\,P^*)$. 

\begin{figure*}
\capstart
\centering
\includegraphics[width = \linewidth]{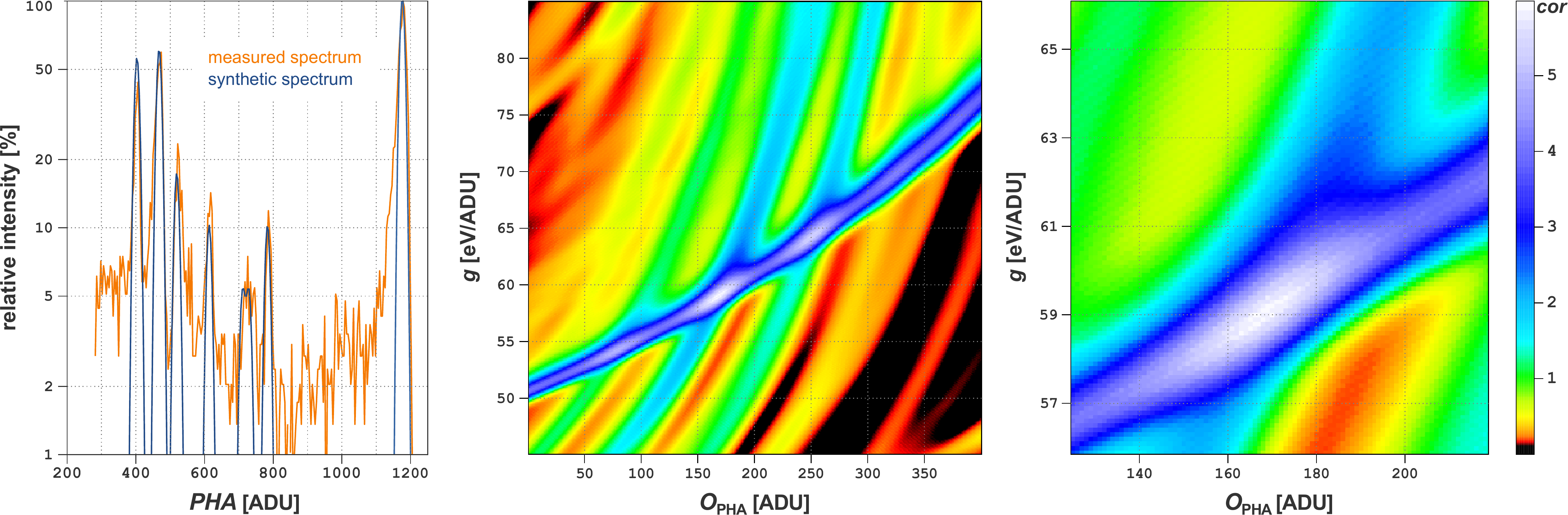}
\caption{Linear ECC shown for one pixel irradiated with \Am.
Left: measured pulse-height spectrum (orange) in comparison to the synthetic spectrum with the highest correlation factor (blue).
Center: color-coded correlation map as a function of the pulse-height offset~$O_\mathrm{PHA}$ and the gain~$g$. Three islands with large correlation factors are visible.
Right: the correlation map near the maximal correlation. The optimal parameters \mbox{$O_\mathrm{PHA} =  167\,$ADU} and \mbox{$g = 59.0$\,eV/ADU} result in a correlation of $C = 6.216$.}
\label{fig:ecc_lin}
\end{figure*}

\section{Application}
\label{sec:app}
The ECC method was tested with a 8\,x\,8 pixel CdTe detector {Caliste~64}~\cite{Meuris2009} within the detector setup CANDELA \cite{Maier2012,Maier2014}. 
The following linear and non-linear application is based on a calibration measurement using the radio-isotope \Am. 

\subsection{Linear calibration}
\label{sec:lin}
The synthetic energy spectrum is constructed using tabulated data \cite{Firestone1999} for the gamma and X-ray emission of \Am~in terms of energy and intensity, see Fig.\,\ref{fig:ECC}a).
The line emissions are convolved with the expected spectral resolution and the quantum efficiency of the detector system which are both considered to be energy dependent, see Fig.\,\ref{fig:ECC}b). 
In the easiest case, a linear model is used to describe the pulse-height-to-energy transformation via
\begin{equation}
   \label{eq:ecc_lin}
	E(\PHA\,|\,{g,\,O_\mathrm{E}}) = g \cdot \PHA + O_\mathrm{E}
  \vspace{-2mm}
\end{equation}
or 
\begin{equation}
	\PHA(E\,|\,{g,\,O_\mathrm{E}}) = g^{-1} ( E - O_\mathrm{E} ) = E/g + O_\mathrm{PHA},
\end{equation}
respectively. Here, the energy offset~$O_\mathrm{E}$ is the energy assigned to \mbox{$\PHA = 0\,\mathrm{ADU}$}, the pulse-height offset~$O_\mathrm{PHA} = -O_\mathrm{E}/g$ is the pulse height assigned to \mbox{$E=0\,$keV}, and the gain~$g$ describes the linear increase of energy per pulse-height channel. Different offset values shift the resulting synthetic pulse-height spectrum to lower or higher pulse-heights, while different gain values squeezes or stretches $I_\mathrm{syn}(\PHA)$, see Fig.\,\ref{fig:ECC}c)-e). 

The correlation factor~$C$ is obtained for different combinations of offset and gain values, which are sampled in equally spaced steps between their upper and lower limits. 

Figure \ref{fig:ecc_lin} shows the color-coded correlation factors for one pixel as a function of $g$ and $O_\mathrm{PHA}$. This correlation map points out that large correlation factors are obtained for different combinations of the parameters $g$ and $O_\mathrm{PHA}$. This effect is caused by the repeating, comb-like structure of the calibration lines of \Am. A short example may illustrate the implication: a combination of (wrong) gain and offset values can result in an alignment of the synthetic Np-L$\beta$, Np-L$\gamma$, and 59\,keV emission lines with the observed Np-L$\alpha$, Np-L$\beta$, and 59\,keV emission lines and result in a large correlation factor. 

Nevertheless, the largest correlation factor is obtained with a correct alignment between synthetic and observed spectrum. The best correlation $C = 6.216$ is obtained for the parameter set $P^* = \{O_\mathrm{PHA} = 167\,\mathrm{ADU}; g = 59.0\,\mathrm{eV/ADU}\}$ and is 87\,\% of $C_\mathrm{max}$, see Eq.\,(\ref{eq:c_max}). 
Figure~\ref{fig:ecc_lin} shows that the corresponding synthetic spectrum $I_\mathrm{syn}(\PHA, P^*)$ matches the observed spectrum very well.
The next local maximum at $P^* = \{O_\mathrm{PHA} = 251\,\mathrm{ADU};\;g = 64.3\,\mathrm{eV/ADU}\}$ with $C =78\,\%\,C_\mathrm{max}$ can be clearly distinguished from the global maximum of $C$.

\subsection{Non-linear calibration}
\label{sec:nl}
\begin{figure}[ht]
\capstart
\centering
\includegraphics[width = \linewidth]{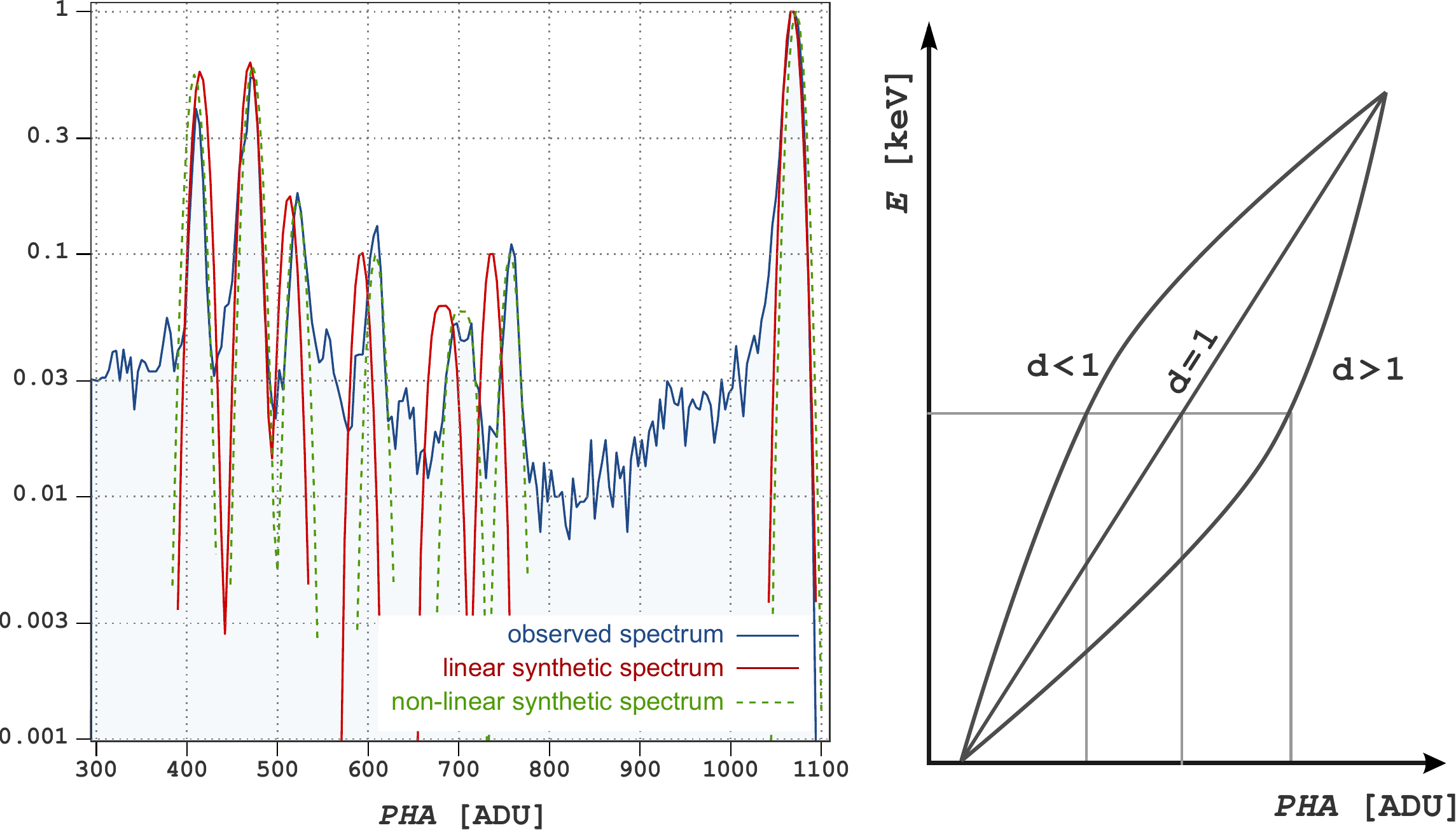}
\caption{Non-linear ECC shown for one pixel irradiated with \Am. Left: the observed spectrum of one pixel superimposed with its best linear and non-linear synthetic spectrum.
Right: the distortion factor $d$ of Eq.\,(\ref{eq:ecc_nl}) shifts the central part of the synthetic spectra to lower ($d<1$) or higher ($d>1$) pulse-height values. $d = 1$ is equivalent to the linear model defined in Eq.\,(\ref{eq:ecc_lin}).}
\label{fig:ecc_nl}
\end{figure}
Some of the pixels of Caliste~64 show a considerable non-linear pulse-height-to-energy relation which is caused by the sensitivity of its front-end readout electronics to the detector dark current which deviates from pixel to pixel. Figure~\ref{fig:ecc_nl} shows the observed spectrum of a pixel with a pronounced non-linear response. The mismatch between the superimposed best synthetic spectrum obtained with a linear model is maximal in the region around $\PHA=700\,\mathrm{ADU}$ and negligible for the dominant emission lines at 470\,ADU and 1070\,ADU, respectively. This reveals an inherent property of the ECC method:  the method tries to optimize the match of the dominant emission lines because they have the largest contribution to the correlation factor in Eq.\,(\ref{eq:cor}). 

The ECC technique allows to correct for the non-linear behavior easily by introducing an appropriate model for the pulse-height-to-energy transformation.
\begin{equation}
  \label{eq:ecc_nl}
	E = g(\PHA-O_\mathrm{PHA})^{d}
\end{equation}
allows to correct for two types of non-linearities with the introduction of a single additional parameter---the non-linearity $d$. For $d<1$ ($d>1$) the central peaks of the synthetic spectra are shifted to lower (higher) PHA values, see the right part of Fig.\,\ref{fig:ecc_nl}.

The three dimensional parameter space $(O_\mathrm{PHA}/g/d)$ results in a correlation cube similar to the correlation map of Fig.\,\ref{fig:ecc_lin}. Figure~\ref{fig:ecc_nl} shows that the optimal non-linear synthetic spectrum matches the observed spectrum very well, i.e.~that Eq.\,(\ref{eq:ecc_nl}) is an adequate model for the energy-to-pulse-height transformation for Caliste~64 in particular, or for only slightly non-linear response systems in general.

\section{Results}
\subsection{Spectroscopic implications}
\label{sec:res}
\begin{figure}[tbp]
\capstart
\centering
\vspace{1.4mm}
\includegraphics[width = \linewidth]{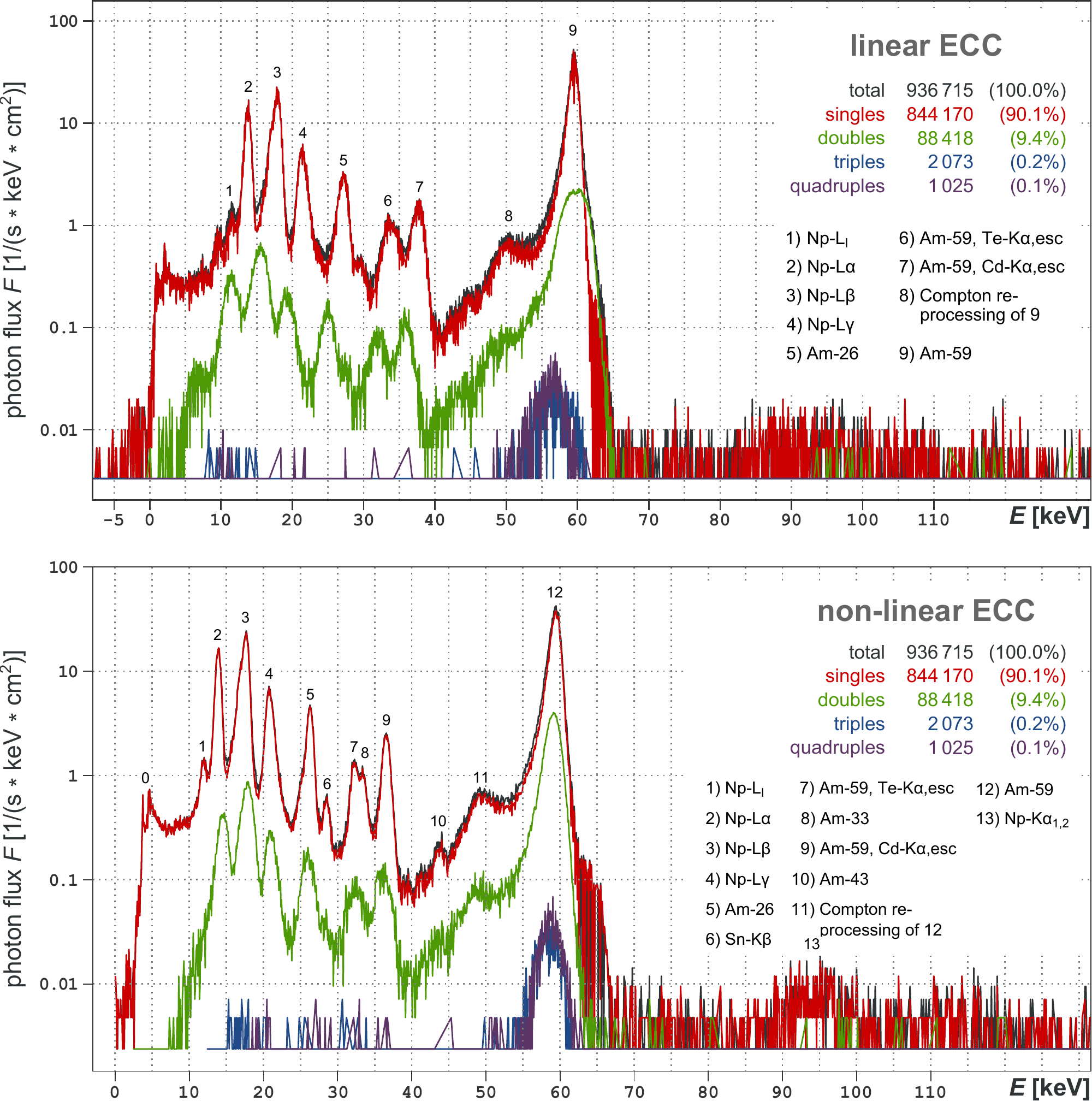}
\caption{\Am~spectra of a 100 minutes observation obtained with a linear ECC (top) and a non-linear ECC (bottom). See also Tab.\,\ref{tab:ecc_error} for more details on the emission lines.}
\label{fig:ecc_spec}
\vspace{-2mm}
\end{figure}
Figure~\ref{fig:ecc_spec} shows the combined \Am~energy spectrum after the pixel specific calibration obtained\footnote{The conditions of the measurement were: temperature $T=10^\circ$C, depletion voltage \mbox{$U = 300\,$V}, shaping time $t = 9.6\,\upmu$s, energy threshold $E_\mathrm{th} = 3\,$keV.} with the linear and the non-linear model. Systematic errors in the energy calibration are investigated through the mismatch between the known energies of the emission lines and the observed energies, see Tab.\,\ref{tab:ecc_error}. The observed energies are obtained via fitting a Gaussian line profile to the observed spectrum using the software GNUPLOT.

Using the non-linear model, the systematic errors are measured to be below $124\,$eV for all listed emission lines. Comparing these systematic uncertainties with the energy of the observed radiation which is in the range of several 10\,keV and with the energy resolution of the used detector which is around 1\,keV demonstrates the quality of the performed calibration.

If the signal charge of an absorbed photon splits into two neighboring pixels, the event is classified as a double event. Because charge loss in the pixel gap can be neglected for Caliste 64 \cite{Meuris2009_cs} the mismatch between the single event spectrum and the double event spectrum indicates calibration errors at energies much lower than the corresponding photon energy. A short example may illustrate this: a split of a Np-L$_\alpha$ photon ($E_0=13.9$\,keV) can result in two events with energies $E_1=10\,$keV and $E_2=3.9\,$keV which are observed at $\PHA_1=336\,$ADU and $\PHA_2=220\,$ADU, respectively. The energy calibration at $\PHA_1$ might be accurate because the energy is close to an EPRP (the Np-L$_\alpha$ line itself). But the energy calibration at $\PHA_2$ can be inaccurate because the energy is much smaller than the lowest energetic EPRP that was used to obtain the calibration parameters. As a result the combined energy of the split can be smaller or larger than the original $E_0$. 

The linear ECC (Fig.\,\ref{fig:ecc_spec}, top) shows strong mismatches between the single event spectrum and the multiple event spectra within the entire observed energy range. For the non-linear ECC (Fig.\,\ref{fig:ecc_spec}, bottom), significant double event mismatches are only observed at 13.9\,keV and 17.8\,keV with a mismatch of 0.56\,keV and 0.28\,keV, respectively. This shows that the non-linear ECC has a better performance for low energies and for multiple events than the linear ECC.

\begin{table*}[t]
\centering
\caption{Summary of the measured systematic errors caused by the energy calibration on single events. The difference between the true line energy $E_0$ and the detected line energy $E$ is calculated via \mbox{$\Delta_{E_0} = E-E_0$} for Np-L$_\upalpha$ ($\Delta_{14}$), Np-L$_\upbeta$ ($\Delta_{18}$), Np-L$_\upgamma$ ($\Delta_{21}$), the 26\,keV emission of \mbox{\Am~($\Delta_{26}$)}, the Cd-K$_\upalpha$ escape peak of the 59\,keV emission of \Am~($\Delta_{36}$), and the 59\,keV emission of \Am~($\Delta_{59}$). The energy resolution (FWHM) at 14\,keV ($\Delta E_1)$ and 59\,keV ($\Delta E_2$) are not affected by the calibration model. The uncertainty of the detected line energy $E$ and of the energy resolutions are also listed in terms of their standard deviations~$\sigma$.\vspace{3mm}}
\begin{tabular}{rcccccccc}
\toprule[0.6mm]
  & $\Delta_{14}$ & $\Delta_{18}$ & $\Delta_{21}$ & $\Delta_{26}$ & $\Delta_{36}$ & $\Delta_{59}$ & $\Delta E_1$ & $\Delta E_2$\\
 \midrule
 lin.\,ECC [eV]  & -138 & 186  & 652 & 759  & 1344   & -84   &  959 & 1024\\
  $\sigma$ [eV] & 16.4 & 26.5 & 18.2 & 25.2 & 28.6 & 36.6 & 84 & 105\\
 \midrule 
non-lin.\,ECC [eV]  & 66   & -73  & 17  & -4  & 124     & -75   &  935 & 1023\\
  $\sigma$ [eV] & 7.9 & 15.0 & 16.1& 13.8 & 13.6 & 18.3 & 93 & 104\\
 \bottomrule[0.6mm]
 \label{tab:ecc_error}
\end{tabular}
\vspace{-5mm}
\end{table*}

\subsection{Performance verification at low counting statistics}
\label{sec:perf}
The enhanced performance of the ECC method compared to a conventional line fitting approach can be demonstrated by performing an energy calibration with both methods for different counting statistics and investigating the calibration errors. In the following, this is shown for a linear energy calibration. 

\subsubsection{Quantifying the calibration error}
In case of the peak fitting approach, the Np-L$\alpha$ and the 59.9\,keV emission line of an \Am~calibration source were used to compute a linear pulse-height-to-energy transformation, i.e.~to compute the gain and the offset. The choice for these two lines is reasonable because they are intensive and spread over a wide energy range. 
The ECC approach was performed as described in Sect.\,\ref{sec:lin} resulting in a gain and an offset value for a specific energy calibration.

To test the accuracy of the energy transformation, i.e.~the accuracy of the obtained gain and offset values, the energy transformation according to Eq.\,(\ref{eq:ecc_lin}) is applied to a central pulse-height value of $\PHA_0=587$\,ADU---the position of the 26\,keV americium line. The absolute difference between the known energy $E_0 = 26\,350\,$eV and the energy value $E$ obtained for this pulse height is considered as the calibration error $\mathit{Err}$ at this central pulse height for the used gain $g$ and offset $O_\mathrm{PHA}$:
\begin{equation}
   \mathit{Err} = \big|E_0 - E(\PHA_0\,|\,g,\,O_\mathrm{PHA}) \big|
\end{equation}
It is worth mentioning that for the different measurements, $\PHA_0$ stayed constant. In this way the uncertainties in the parameters $g$ and $O_\mathrm{PHA}$ can be evaluated at 26\,keV without adding the uncertainty of finding the peak position of the 26\,keV americium emission line.

\subsubsection{Analysis}
\begin{figure}[ht]
   \capstart
   \centering
   \vspace{0.5mm}
	 \includegraphics[width = \linewidth]{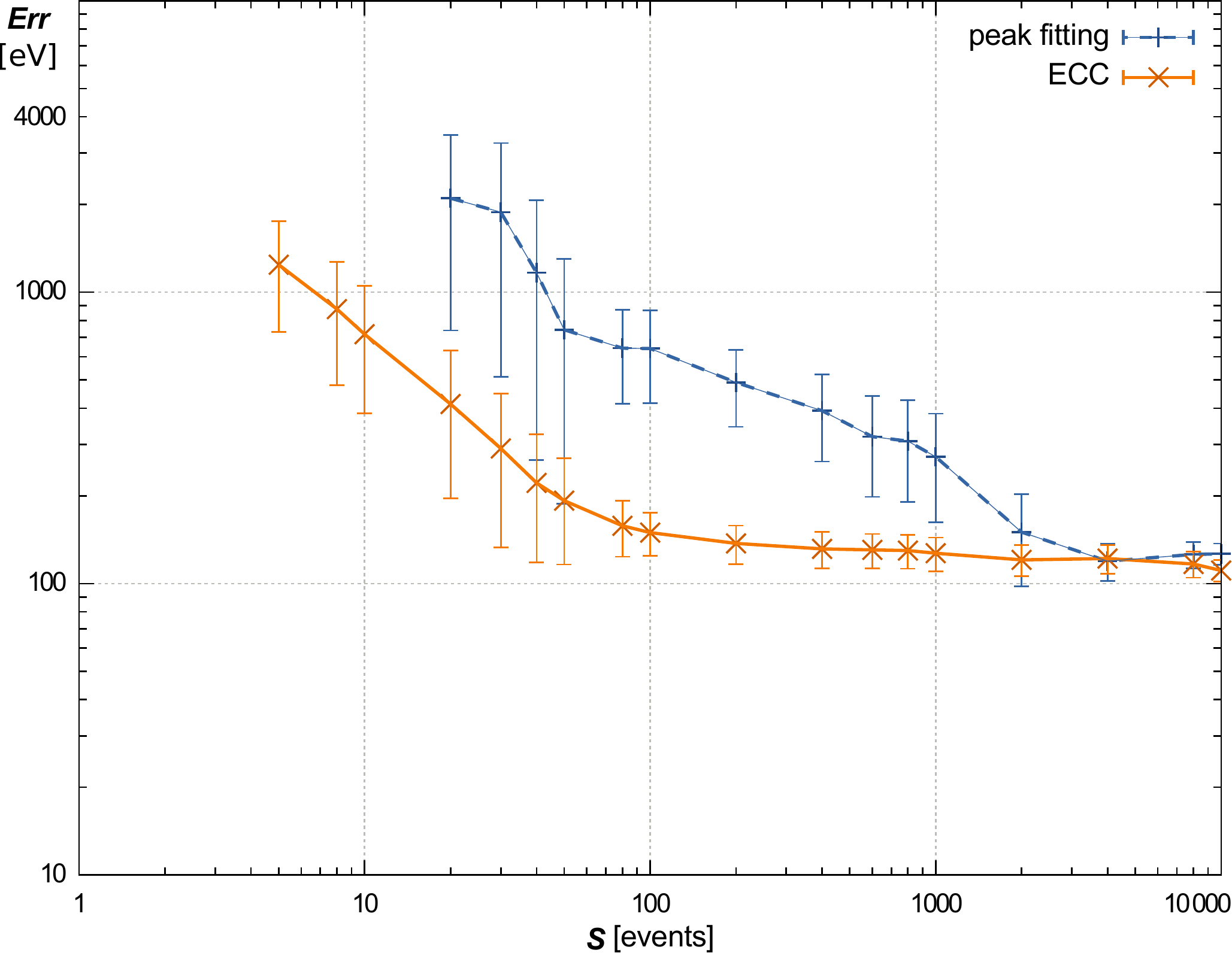}
   \caption{Absolute calibration error $\mathit{Err}$ as a function of the sample size $S$ estimated at an energy of 26.35\,keV for a calibration based on peak fitting or on ECC, respectively. The break down of the ECC calibration occurs at a sample size witch is approximately a factor of ten lower compared to a calibration which is based on peak fitting.}
   \label{fig:performance}
	\vspace{-3mm}
\end{figure}

The calibration error $\mathit{Err}$ calculated for different sampling sizes $S$ varies strongly for small values of $S$. In order to eliminate statistical fluctuations of the observed calibration errors, multiple measurements for a specific sample size are analyzed. In fact, all the different samples are produced on the basis of one observation with 16\,000 events observed within one pixel of Caliste 64. This way, systematical effects which originate from different data sets can be neglected. Taking data of the intervals [1..S], [2..S+1], ..., [16\,000-S+1..16\,000] results in $16\,000 - S + 1$ different spectra and the same amount of gains, offsets, and calibration errors. The mean value and the standard deviation of $\mathit{Err}$ are shown in Fig.\,\ref{fig:performance} as a function of the sample size $S$. Because the frequency distribution of $\mathit{Err}$ can only be assumed to be Gaussian-like near its mean value the error bars are plotted in a small interval of $\sigma/3$. 

Even though an accurate estimation of the true, non-Gaussian interval estimations would require a more fundamental approach than the simple calculation of the standard deviation two important conclusions can be made:
\begin{itemize}
\item for large sample sizes ($S > 4000$\,events) peak fitting and ECC result in the same, constant calibration error which can be explained by the non-linearity of the used detector system.
\item for small sample sizes both methods result in increased mean calibration errors and in increased variations around these mean values. But, this break down of the energy calibration occurs for the ECC approach at sample sizes which are an order of magnitude smaller than in the case of the peak fitting approach. The calibration error with ECC at $S_\mathrm{ECC}=100$\,events is comparable with the calibration error obtained with a peak fitting approach at $S_\mathrm{fit}=2000\,$events. These two sample sizes, $S_\mathrm{ECC}$ and $S_\mathrm{fit}$, define the limits of the two methods. At lower counting statistics the calibration error increases rapidly.
\end{itemize}

The presented peak fitting method could be improved by using more emission lines than just the two presented ones. The increased uncertainty of the peak position of the fainter lines could be compensated with a weighting factor for each line according to its strength.  This modified peak fitting method should approximate the calibration errors obtained with the ECC method in case of an emission-line dominated calibration spectrum. In fact, the two methods are very similar in this scenario, although the ECC method is easier to apply.

\section{Discussion}
\label{sec:disc}
Even though this work concentrates on the description of the basic principle of the ECC method, a few considerations on its statistical performance and on the possible calibration sources are presented in the following. 
The core of the ECC method is to find the optimal calibration model by varying the parameters of the model. This is indeed similar to a fitting approach; the
main difference between both methods is that the ECC method uses all PHA
values---and not just the peak positions. In this way, more information from the
calibration measurement is used, which results in an enhanced statistics of the
ECC technique compared to a line fitting approach. 
The shape of the calibration spectrum plays an essential role for the estimation of the calibration performance. Besides line emission spectra, continuous spectra originating from X-ray tubes or synchrotron beams can be used for calibration. 

\subsection{Line spectra} 
Synthetic spectra based on radio-isotopes and/or X-ray fluorescence can be constructed using tabulated nuclear and atomic data. See \cite{Cho2014} for a detailed study of the generation of X-ray fluorescence for energy calibration. The number of emission lines should be as many as possible and they should be distributed homogeneously over the energy range of the detector. Additionally, the autocorrelation of the source spectrum should have a clear maximum in order to avoid the detection of erroneous sidelobes during calibration.

If the spectrum has a very dominant emission line the calibration via correlation can be erroneous because all parameters are optimized in order to match the shape of this line: an inaccurate estimation of the line width is then, for example, compensated with a wrong parameter set. This effect can be minimized with a proper selection of the calibration sources or with an artificial attenuation (by soft- or hardware) of the dominant emission line.

Besides the already mentioned benefit of using the whole spectral shape and not just the peak position of an emission line, the ECC method has another advantage compared to a line fitting approach. Emission lines which cannot be resolved by the detector system result in a broad emission hump. This hump can be included in the synthetic spectrum and can be used for calibration; this feature is especially beneficial for detector systems with a lower spectroscopic resolution than the presented one.

\subsection{Continuous spectra}
By construction, the ECC method works optimal with a continuous calibration spectrum. The continuous synthetic energy spectrum can be based on theoretical models, on simulations---see \cite{Youn2014} for simulations of X-ray tube spectra---or on measurements with a calibrated spectrometer. 

In principle, the calibrated spectrometer can be the detector itself being in a calibrated state. This way, changes in the detector response can be observed by monitoring the correlation between the calibrated pulse-height spectrum and the actual pulse-height spectrum. If this correlation falls below a critical threshold, a recalibration can be done using the same data set.

\subsection{Extending the parameter space}
In order to conclude from the measured spectral flux to the real spectral flux, a precise knowledge of the detector efficiency is necessary. In the presented work, the quantum efficiency $\mathit{Q}$ was calculated on the basis of the detector material (with density $\rho$), the detector thickness $d$, and the (energy dependent) cross section for photoelectric absorption $\sigma(E)$ according to 
\begin{equation}
   \mathit{Q}(E) = \rho \cdot d \cdot \sigma(E).
\end{equation}
A more detailed and pixel specific quantum efficiency can be obtained with the ECC method by introducing a parametrized model for the quantum efficiency and adding these parameters to the parameter set $P$. 

In a similar way, the energy resolution of the detector system can be obtained pixel specific and as a function of energy. It should be mentioned that the computation time for the correlation increases linearly with the number of elements in the parameter space.

\subsection{Calculation performance}
The performance of the ECC calculations can be increased with a parameter sampling which is not equally spaced but which samples the total parameter space first roughly and becomes then more and more detailed near the maximum of the correlation. For the presented non-linear calibration which uses 400 different offset values, 400 gain values, and 25 distortion factors, the computation time for 64 pixels with an average personal computer was in the range of a few minutes. The potential to increase the calculation performance seems promising, as the code is not optimized in terms of performance, the sampling is equally spaced, and the algorithm can be parallelized to a large degree.

\section{Conclusion} 
Energy calibration via correlation can be used with all kinds of calibration sources---radioactive isotopes, X-ray tubes, or synchrotron
beams. A set of emission lines originating from radioactive isotopes in combination with X-ray fluorescence may serve for many applications the most easy and flexible solution. With such a setup, a calibration of the 14-60\,keV energy band was performed with a remaining systematic error less than $~\!\!0.1\,$keV. 

A broad band source emission uses the benefits of the method best. Besides the presented cases of a linear and a non-linear energy calibration the ECC method can be used for any kind of functional relation between pulse-height values and energies. Furthermore, even though the concept of calibration via correlation is presented in the context of energy calibration it can be applied to any kind of calibration which is based on the comparison of a measured quantity to a parametric model.

\bibliography{lit}
\bibliographystyle{abbrvnat}

\end{document}